# Accounting for the 'network' in the Natura 2000 network: A response to *Hochkirch et al.* 2013

Jan O. Engler[1,2*], Anna F. Cord[3], Petra Dieker[4,5], J. Wolfgang Wägele[1] & Dennis Rödder[1]

[1]*Zoological Researchmuseum Alexander Koenig, Adenauerallee 160, D-53113 Bonn, Germany*

[2]*Department of Wildlife Sciences, University of Göttingen, Büsgenweg 3, D-37077 Göttingen, Germany*

[3]*Helmholtz Centre for Environmental Research – UFZ, Department of Computational Landscape Ecology, Permoserstraße 15, D-04318 Leipzig, Germany*

[4]*Helmholtz Centre for Environmental Research – UFZ, Department of Community Ecology, Theodor-Lieser-Straße 4, D-06114 Halle, Germany*

[5]*Centre for Methods, Institute of Ecology, Leuphana University Lüneburg, Scharnhorststraße 1, D-21335 Lüneburg, Germany*

*corresponding Author E-mail: j.engler.zfmk@uni-bonn.de

Worldwide, we are experiencing an unprecedented, accelerated loss of biodiversity triggered by a bundle of anthropogenic threats such as habitat destruction, environmental pollution and climate change (Butchart et al. 2010). Despite all efforts of the European biodiversity conservation policy – initiated 20 years ago by the Habitats Directive (EU 1992) that provided the legal basis for establishing the Natura 2000 network – the goal to halt the decline of biodiversity in Europe by 2010 has been missed (EEA 2010). Hochkirch et al. (2013) identified four major shortcomings of the current implementation of the directive concerning prioritization of the annexes, conservation plans, survey systems and financial resources. They hence proposed respective adaption strategies for a new Natura 2020 network to reach the Aichi Biodiversity Targets.

Despite the significance of these four aspects, Hochkirch et al. (2013) did not account for the intended 'network' character of the Natura 2000 sites, an aspect of highest relevance. Per definition, a network requires connective elements (i.e. corridors) between its nodes. From an ecological perspective, the Natura 2000 network must guarantee that the species of concern are able to exchange between habitat patches (above all for maintaining/fostering gene flow; e.g. Storfer et al. 2007). Several studies have shown that reserves fail to protect the species they were designed for due to their isolated character in an anthropogenically degraded landscape matrix (e.g. Seiferling et al. 2012), even though they are well managed (Filz et al. 2013). In turn, habitat connectivity greatly enhances the movement of species within fragmented landscapes (Gilbert-Norton et al. 2010). Both Habitats (Art. 10) and Birds Directive (Art. 3) explicitly mention the importance of elements providing functional connectivity ('ecological coherence') outside the designated Special Areas of Conservation (SACs) for species of Community interest. However, since the member states are responsible for the designation of SACs, their selection often represents a consensus of various political, economic and ecological considerations. This weakness is well acknowledged in a guidance document from the Institute for European Environmental Policy (Kettunen et al. 2007). The authors formulated a framework for assessing, planning and implementing ecological connectivity measures in a way that is legally binding and standardized across borders. Additionally, they presented measures increasing habitat connectivity and future research needed on this topic. Besides the strategies proposed by Hochkirch et al. (2013), there is hence an urgent need to investigate the inter-reserve connectivity in the Natura 2000 network as a whole and specifically for the priority species for which SACs have been designated. Recent software developments and the increasing availability of high-resolution environmental data in combination with extensive fieldwork will help to meet these research requirements. Finally, the results derived from such research must be implemented into a binding EU-legislation as well as a standardized planning policy across national borders to reach scientific consensus on corridor design, which often lacked in the past (Bennett et al. 2006). This might ultimately ensure an ecological coherence between SACs, which is the prerequisite, over any other strategies, ensuring a Natura 2020 network being worth its name.